\documentclass[conference]{IEEEtran}
\IEEEoverridecommandlockouts
\usepackage{cite}
\usepackage{amsmath,amssymb,amsfonts,bbm}
\newcommand{\norm}[1]{\left\lVert#1\right\rVert}
\usepackage{algorithmic}
\usepackage{graphicx}
\usepackage{subfigure}
\usepackage{textcomp}
\usepackage{xcolor}
\usepackage{booktabs}
\newcommand{\ra}[1]{\renewcommand{\arraystretch}{#1}}

\graphicspath{{./img/}}

\ifodd 2
\newcommand{\revQZ}[1]{{\color{blue}#1}} 
\newcommand{\comQZ}[1]{\textbf{\color{red}(COMMENT [QZ]: #1)}}
\newcommand{\comXT}[1]{\textbf{\color{red}(COMMENT [XT]: #1)}}
\else
\newcommand{\revQZ}[1]{#1}

\newcommand{\comQZ}[1]{}
\newcommand{\comXT}[1]{}
\fi

\def\BibTeX{{\rm B\kern-.05em{\sc i\kern-.025em b}\kern-.08em
    T\kern-.1667em\lower.7ex\hbox{E}\kern-.125emX}}
\begin{document}

\title{ECG Reconstruction via PPG: A Pilot Study}

\author{\IEEEauthorblockN{Qiang~Zhu\IEEEauthorrefmark{1}\thanks{\IEEEauthorrefmark{1}\{zhuqiang, xtian17, minwu\}@umd.edu, \IEEEauthorrefmark{2}chauwai.wong@ncsu.edu.},~\IEEEmembership{Student Member,~IEEE,}
		Xin Tian\IEEEauthorrefmark{1},~\IEEEmembership{Student Member,~IEEE,}
		Chau-Wai~Wong\IEEEauthorrefmark{2},~\IEEEmembership{Member,~IEEE,}\\
        and~Min~Wu\IEEEauthorrefmark{1}, ~\IEEEmembership{Fellow,~IEEE}\\}
 \IEEEauthorblockA{\IEEEauthorrefmark{1}Department of Electrical and Computer Engineering, University of Maryland, College Park, USA\\ \IEEEauthorrefmark{2}Department of Electrical and Computer Engineering, North Carolina State University, Raleigh, USA}}

\maketitle

\begin{abstract}
In this paper, the relation between electrocardiogram (ECG) and photoplethysmogram (PPG) signals is studied, and the waveform of ECG is inferred via the PPG signals. \revQZ{In} order to address this inverse problem, a transform is proposed to map the discrete cosine transform (DCT) coefficients of each PPG cycle to those of the corresponding ECG cycle. The resulting DCT coefficients of the ECG cycle are inversely transformed to obtain the reconstructed ECG waveform. \revQZ{The proposed method is evaluated} on a benchmark dataset of subjects with a variety of combinations of age and weight. Experimental results show that \revQZ{the proposed method} can achieve \revQZ{a} high accuracy at 0.98 in averaged correlation.
\end{abstract}

\begin{IEEEkeywords}
ECG, PPG, inverse problem, DCT.
\end{IEEEkeywords}

\section{Introduction}\label{sec::intro}
\par The electrocardiogram (ECG) has become the most commonly used cardiovascular diagnostic procedure and is a fundamental tool of clinical practice~\cite{kligfield2007recommendations }\comQZ{,fye1994history}. Many modern wearable ECG systems have been developed in recent decades. They are simpler and more reliable than before, weighing only a fraction of a pound. However, the material used to provide good signal quality with the electrode may cause skin irritation and discomfort during prolonged use, which restricts the long-term use of the devices.

The photoplethysmogram (PPG) is a noninvasive circulatory signal related to the pulsatile volume of blood in tissues~\cite{reisner2008PPGUtility}. Compared with ECG, PPG is easier to set up, more convenient, and more economical. PPG is nearly ubiquitous in clinics and hospitals in the form of finger/toe clips and oximeters and has increasing popularity in the form of consumer-grade wearable devices that offer continuous and long-term monitoring capability and do not cause skin irritations. 
 
The PPG and ECG signals are intrinsically correlated, considering that the variation of the peripheral blood volume is influenced by the left ventricular myocardial activities, and these activities are controlled by the electrical signals originating from the sinoatrial (SA) node. The timing, amplitude, and shape characteristics of the PPG waveform contain information about the interaction between the heart and the connective vasculature. These features have been translated to measure heart rate, heart rate variability, respiration rate~\cite{karlen2013CapnobaseRR}\comQZ{nilsson2005respiration}, blood oxygen saturation~\cite{aoyagi2002pulseoximetry}, blood pressure~\cite{payne2006PTTBloodPressure}, and to assess vascular function~\cite{marston2002ppgChronicVenousInsufficiency,allen1993PPG4PAD}.\comQZ{ allen2000PPG4PAD} As the prevailing use of wearable device capturing users' PPG signal on a daily basis, we are inspired to utilize this correlation to not only infer the ECG parameters but also reconstruct the ECG waveform from the PPG measurement. This exploration, if successful, can provide a low-cost ECG screening for continuous and long-term monitoring and take advantage of both the \revQZ{rich clinical knowledge base} of ECG signal and the easy accessibility of the PPG signal.

There is a very limited amount of prior art addressing the ECG reconstruction/\revQZ{inference} problem mentioned above. In~\cite{banerjee2014photoecg}, the authors trained several classifiers to infer the quantized level of RR, PR, QRS, and QT interval parameters, respectively, from selected time domain and frequency domain features of PPG. Even though the system yields $90\%$ accuracy on a benchmark hospital dataset, the capability confined to only inferring ECG parameters may restrict \comQZ{its clinical use.}\revQZ{the broad adoption of this prior work.}

In this paper, we propose to estimate the waveform of the ECG signal using PPG measurement by learning a signal model that relates the two time series. We first preprocess the ECG and PPG signal pairs to obtain temporally aligned and normalized sets of signals. We then segment the signals into pairs of cycles and train a linear transform that maps the discrete cosine transform (DCT) coefficients of the PPG cycle to those of the corresponding ECG cycle. The ECG waveform is then obtained via the inverse DCT. 

The significance of this work is threefold. First, the statistics of the system performance metrics \revQZ{evaluated on a benchmark database} show that our proposed system can reconstruct the ECG signal accurately. 
Second, to the best of our knowledge, this is the first work which addresses the problem of \revQZ{inferring} ECG waveform from the PPG signal. It may open up a new direction for cardiac medical practitioners, wearable technologists, and data scientists to leverage a rich body of clinical ECG knowledge and transfer the understanding to build a knowledge base for PPG and data from wearable devices. 
Third, the technology may enable a more user-friendly, low-cost, continuous and long-term cardiac monitoring that supports and promotes public health, especially for people with special needs.

\begin{figure*}
\centering
\includegraphics[width=7in]{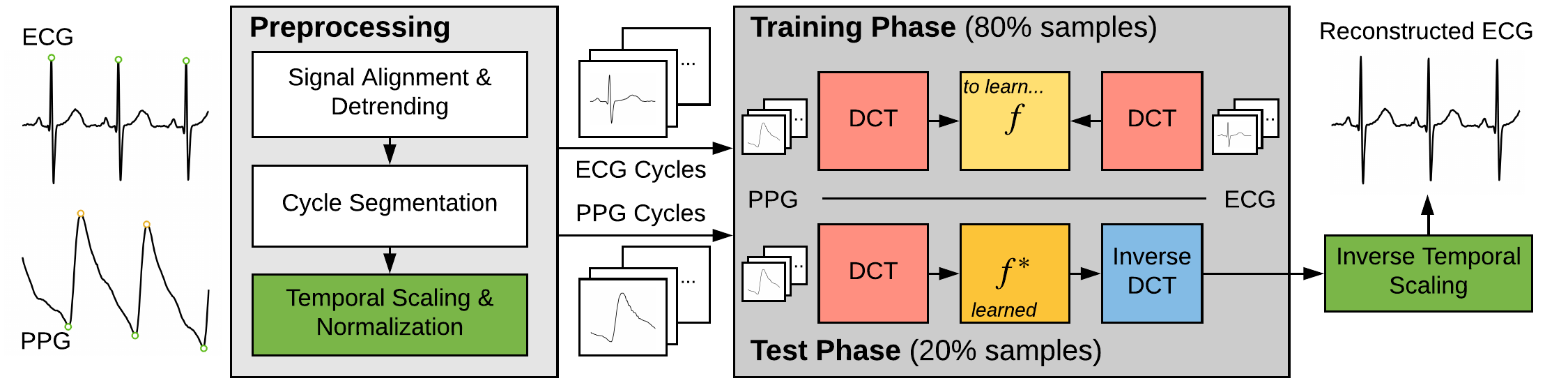}
\caption{Flowchart of the proposed system. The ECG and PPG signals are first preprocessed to obtain physically aligned and normalized pairs of cycles. The selected DCT coefficients of $80\%$ pairs of cycles are used for training a linear transform $f$ which is used in the test phase to reconstruct the ECG signals.}
\label{fig::sysdiag}
\end{figure*}

\section{\revQZ{Proposed System}}\label{sec::sys}
\subsection{Preprocessing: Cycle-Wise Segmentation}\label{subsec::sys_preproc}
The goal of preprocessing ECG and PPG signals is to obtain temporally aligned and normalized pair of signals, so that the critical temporal features of both waveforms are synchronized to facilitate our investigation. The preprocessing phase shown in Fig.~\ref{fig::sysdiag} contains data alignment, signal detrending, cycle-wise segmentation, temporal scaling, and normalization that be explained as follows. 

\paragraph{Data alignment} Considering possible misalignment of the signal pair in each trial, we perform a two-level signal alignment to obtain physically aligned signal pairs. We first estimate the signal delay in the cycle level using the peak features as they are \revQZ{most distinguishable} within the cycle. We then align the signals to the sample level based on their physical correspondence. 

Consider a pair of almost simultaneously recorded PPG and ECG signals, denoted as $\mathbf{x}\in \mathbb{R}^T$ and $\mathbf{y}\in \mathbb{R}^T$ respectively. We name the coordinate of the systolic peak in the $i$th cycle of PPG as $t_{\text{sp}}(i)$ and the R peak of ECG as $t_{\text{rp}}(i)$. The cycle delay $n_{\text{delay}}$ is \revQZ{estimated from a candidate set} $\mathbb{D}\triangleq[-k, k]$, where the search radius $k=5$ as we expect the cycle delay to be small. For each evaluated $n\in \mathbb{D}$, we first preliminarily align the signal with respect to $t_{\text{sp}}(1-n\cdot\mathbbm{1}(n<0))$, and $t_{\text{rp}}(1-n\cdot\mathbbm{1}(n>0))$. The aligned coordinates of PPG and ECG peaks are $\{t'_{\text{sp}}(n)\}$ and $\{t'_{\text{rp}}(n)\}$. We then estimate the cycle delay $\hat{n}_{\text{delay}}$ by solving the following problem:
\begin{equation}
\begin{split}
    \hat{n}_{\text{delay}}=\underset{n\in\mathbb{D}}{\mathrm{argmin}}\sum_{i=1}^{i=N-k}&\big|t'_{\text{sp}}(i-n\cdot\mathbbm{1}(n<0))\\ &\quad -t'_{\text{rp}}(i+n\cdot\mathbbm{1}(n>0))\big|,
\end{split}\label{eq::SigCycleAlign}
\end{equation}
where $N$ is total number of cycles, $\mathbbm{1}$ is the indicator function. We align the signals by shifting PPG signal so that the systolic peaks of PPG and the R peaks of ECG are temporally matched. 

Next, we align the signal to the sample level according to the R peak of the ECG and the onset point of PPG in the same cycle (namely, the local minimum point before the systolic peak), considering that the R peak corresponds approximately to the opening of the aortic valve, and the onset point of PPG indicates the arrival of the pulse wave~\cite{reisner2008PPGUtility}. In this way, we eliminate the pulse transit time and align the signals. \revQZ{Note that our signal model assumes the PPG and ECG cycles being accurately estimated. In practice, a degradation of the system performance is possible when the signal cycles are estimated inaccurately due to the presence of signal artifacts or pathological disturbances.}
\paragraph{Detrending} The non-stationary trend in both signals can be problematic for temporal pattern analysis. Such slowing-varying trend can be estimated and then subtracted from the original signals. The trend is assumed to be a smooth, unknown version of $\mathbf{x}$ and $\mathbf{y}$ with a property that its accumulated convexity measured for every point on the signal is as small as possible, namely,
\begin{equation}
{\hat{\mathbf{x}}_{\text{trend}}}=\underset{\hat{\mathbf{x}}}{\mathrm{argmin}}\norm{\mathbf{x}-\hat{\mathbf{x}}}_2^2+\lambda\norm{\mathbf{D}_2\hat{\mathbf{x}}}_2^2,
\label{eq::detrend}
\end{equation}
where $\mathbf{x}$ is the original signal, $\hat{\mathbf{x}}_{\text{trend}}$ is the estimated trend in $\mathbf{x}$, $\lambda$ is a regularization parameter controlling the smoothness of the estimated trend, and $\mathbf{D}_2 \in \mathbb{R}^{T\times T}$ is a Toeplitz matrix that acts as a second-order difference operator. The closed-form solution of~\eqref{eq::detrend} is $\hat{\mathbf{x}}_{\text{trend}} = (\mathbf{I} + \lambda \mathbf{D}_2^\intercal \mathbf{D}_2)^{-1}\mathbf{x}$, where $\mathbf{I}$ is the identity matrix, Hence, the detrended signal is $\Tilde{\mathbf{x}}=\mathbf{x}-\hat{\mathbf{x}}_{\text{trend}}$, and similarly, $\Tilde{\mathbf{y}}=\mathbf{y}-\hat{\mathbf{y}}_{\text{trend}}$.
\paragraph{Segmentation $\&$ Normalization}
After the signal alignment and detrending, we segment each cycle of the signal $\Tilde{\mathbf{x}}$ and $\Tilde{\mathbf{y}}$ to prepare for the learning phase. In our experiment, we introduce the following two cycle segmentation schemes:
\begin{itemize}
    \item \textit{SR}: we segment the signal according to the points which are $1/3$ of the cycle length to the left of the R peaks of the ECG signal. We call this scheme SR as it approximately captures the standard shape of sinus rhythm.
    \item \textit{R2R}: we segment the signal according to the location of the R peak of the ECG signal to mitigate the reconstruction error in the QRS complex.
\end{itemize}
After the segmentation, we temporally scale each cycle sample via linear interpolation to make it of length $L$ in order to mitigate the influence of the heart rate variation. We then normalize each cycle by subtracting the sample mean and dividing by the sample standard deviation. We denote the normalized PPG and ECG cycle samples as $\mathbf{C}_x$, $\mathbf{C}_y\in \mathbb{R}^{N\times L }$.

\subsection{Learning a Linear Transform for DCT Coeffients}\label{subsec::sys_learning}
DCT has been shown in the literature to have competitive performance in compressing and representing PPG
and ECG signals~\cite{gholamhosseini1998ECGBasisCompare}. In this study, we use DCT coefficients to compactly represent the ECG and PPG signals. In the training phase, we build and train a linear transform to model the relation between the DCT coefficients of PPG and ECG cycles. We then use the trained matrix to reconstruct the ECG waveform in the test phase.

Specifically, we first perform cycle-wise DCT on $\mathbf{C}_x$ and $\mathbf{C}_y$, which yields $\mathbf{X}$, $\mathbf{Y}\in \mathbb{R}^{N\times L }$. Then the first $L_x, \ L_y$ DCT coefficients of $\mathbf{X},\mathbf{Y}$ are selected to represent the corresponding waveform as the signal energy is concentrated mostly on the lower frequency components per our observation. We denote them as $\Tilde{\mathbf{X}}\in \mathbb{R}^{N\times L_x }$ and $\Tilde{\mathbf{Y}}\in \mathbb{R}^{N\times L_y }$. We next separate $\Tilde{\mathbf{X}}$ and $\Tilde{\mathbf{Y}}$ into training and test sets as $\mathbf{X}_{\text{train}}\in \mathbb{R}^{N_{\text{train}}\times L_x}$, $\mathbf{Y}_{\text{train}}\in \mathbb{R}^{N_{\text{train}}\times L_y}$ and $\mathbf{X}_{\text{test}}\in \mathbb{R}^{N_{\text{test}}\times L_x }$,  $\mathbf{Y}_{\text{test}}\in \mathbb{R}^{N_{\text{test}}\times L_y}$, where $N_{\text{train}}+N_{\text{test}}=N$.

In the training process, a linear transform matrix $f^*\in \mathbb{R}^{L_x\times L_y}$ that maps from PPG to ECG DCT coefficients is learned through ridge regression as described below:
\begin{equation}
    f^* = \underset{f}{\mathrm{argmin}}\norm{ \mathbf{X}_{\text{train}}\  f-\mathbf{Y}_{\text{train}}}_\text{F}^2+\gamma\norm{f}_\text{F}^2,
    \label{eq::ftrianing}
\end{equation}
where $\norm{\cdot}_\text{F}$ denotes the Frobenius norm of a matrix, and $\gamma>0$ is a complexity parameter that controls the shrinkage of $f$ toward zero. The penalization the sum-of-squares of $f$ is to reduce the variance of the predictions and to avoid overfitting~\cite{friedman2001elementsStatLearning}. The analytic solution to~\eqref{eq::ftrianing} is $ f^{*} = (\mathbf{X}_{\text{train}}^\intercal\mathbf{X}_{\text{train}}+\gamma \mathbf{I})^{-1}\mathbf{X}_{\text{train}}^\intercal\mathbf{Y}_{\text{train}}$, where $\mathbf{I}$ is the identity matrix.

In the test phase, we apply the optimal linear transform $f^*$ learned in training stage on $\mathbf{X}_\text{test}$ and estimate the corresponding DCT coefficients of ECG cycles. We denote the estimate as $\hat{\Tilde{\mathbf{Y}}}_\text{test}\triangleq \mathbf{X}_\text{test}\ f^*$. To reconstruct ECG, we first augment each row of $\hat{\Tilde{\mathbf{Y}}}_\text{test}$ to be in the same dimension as $L$ (by padding zeros). We denote the zero-padded matrix as $\hat{\mathbf{Y}}_\text{test} \in \mathbb{R}^{N_\text{test}\times L}$. We then apply inverse DCT to each row of $\hat{\mathbf{Y}}_\text{test}$ and concatenate the resulted temporal matrix row by row to obtain the reconstructed ECG signal $\hat{\mathbf{y}}_\text{test}$.

\begin{figure}[!t]
\centerline{\includegraphics[width=0.45\textwidth]{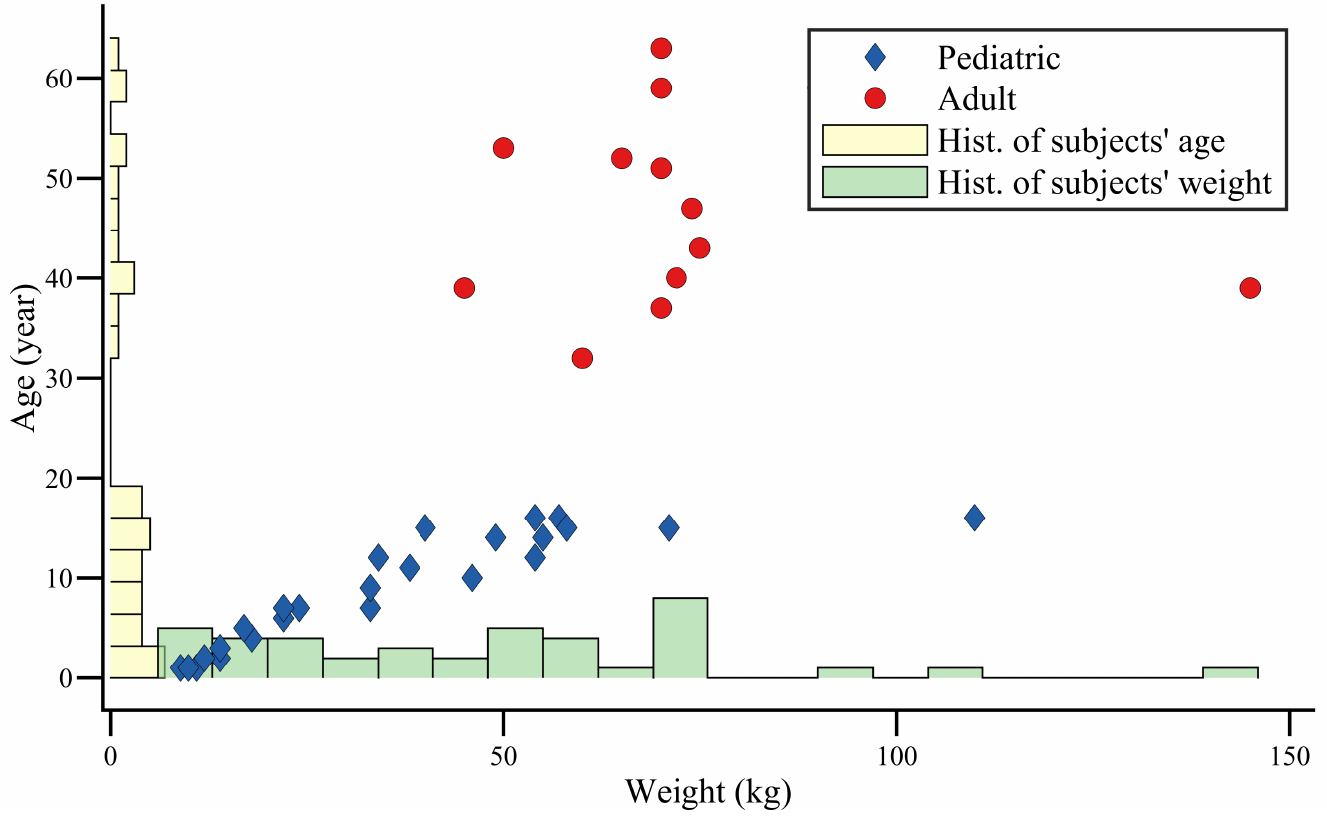}}
\caption{Scatter plot of the age vs. weight of all the subjects in the database~\cite{karlen2013CapnobaseRR}. The bar plot on each x and y axis shows the histogram of subject's weight and age, respectively.}
\label{fig:CapnoDist}
\end{figure}

\section{Experiment Results}\label{sec::result}
We use the Capnobase TBME-RR~\cite{karlen2013CapnobaseRR} to evaluate the performance of the proposed system. The dataset contains 42 eight-min sessions of simultaneously recorded PPG and ECG measurements from $29$ pediatric surgeries and $13$ adult surgeries\revQZ{\footnote{Note that the recording in this database is of high signal quality. In cases when the signal is corrupted by noise or subject's motion artifacts, a denoising process is \revQZ{needed to clean} the signal before the preprocessing stage.}}, sampled at $300$ Hz. Each session corresponds to a unique subject. The PPG signal was acquired on subjects' fingertips via a pulse oximeter. As shown in Fig.~\ref{fig:CapnoDist}, the dataset has a wide variety of patient's age and weight and is thus an ideal dataset for testing the performance of our system.
\par We first pruned the signals according to the human-labeled artifact segments and processed the pairs of ECG and PPG signal using the method introduced in Section~\ref{subsec::sys_preproc} to \revQZ{obtain} aligned and normalized pairs of the signal cycles. We set $L=300$, \revQZ{and} $L_y=100$, as most of the diagnostic information of ECG is contained below $100$~Hz~\cite{kligfield2007recommendations}. We set $\lambda=500$, and $\gamma=10$ empirically as they offer the best regularization results in the tasks. In order to test the consistency of the system, we selected the first $80\%$ of each session as the training set and the rest for testing. In this study, we evaluate the system in a subject-dependent fashion, which means that the linear transform $f^*$ is trained and tested individually in each session. We use the following two metrics to evaluate the system performance in the test set:
\begin{itemize}
    \item Relative root mean squared error: 
    \begin{equation}
        \text{r}\textsc{Rmse}=\frac{\norm{\mathbf{y}_\text{test}-\hat{\mathbf{y}}_\text{test}}_2}{\norm{\mathbf{y}_\text{test}}_2},
    \end{equation}
    
    \item Pearson's correlation coefficient:
    \begin{equation}
      \rho=\frac{(\mathbf{y}_\text{test}-\bar{y}_\text{test})^\intercal(\hat{\mathbf{y}}_\text{test}-\bar{\hat{y}}_\text{test})}{\norm{\mathbf{y}_\text{test}-\bar{y}_\text{test}}_2 \norm{\hat{\mathbf{y}}_\text{test}-\bar{\hat{y}}_\text{test}}_2},  
    \end{equation}
    
\end{itemize}
where $\mathbf{y}_\text{test}$, $\bar{\hat{y}}_\text{test}$, and $\bar{y}_\text{test}$ denote the ECG signal in test set, the average of all coordinates of the vectors $\hat{\mathbf{y}}_\text{test}$ and $\mathbf{y}_\text{test}$ respectively.

\begin{figure}[!t]
\centering
\subfigure[]{\includegraphics[width=0.24\textwidth]{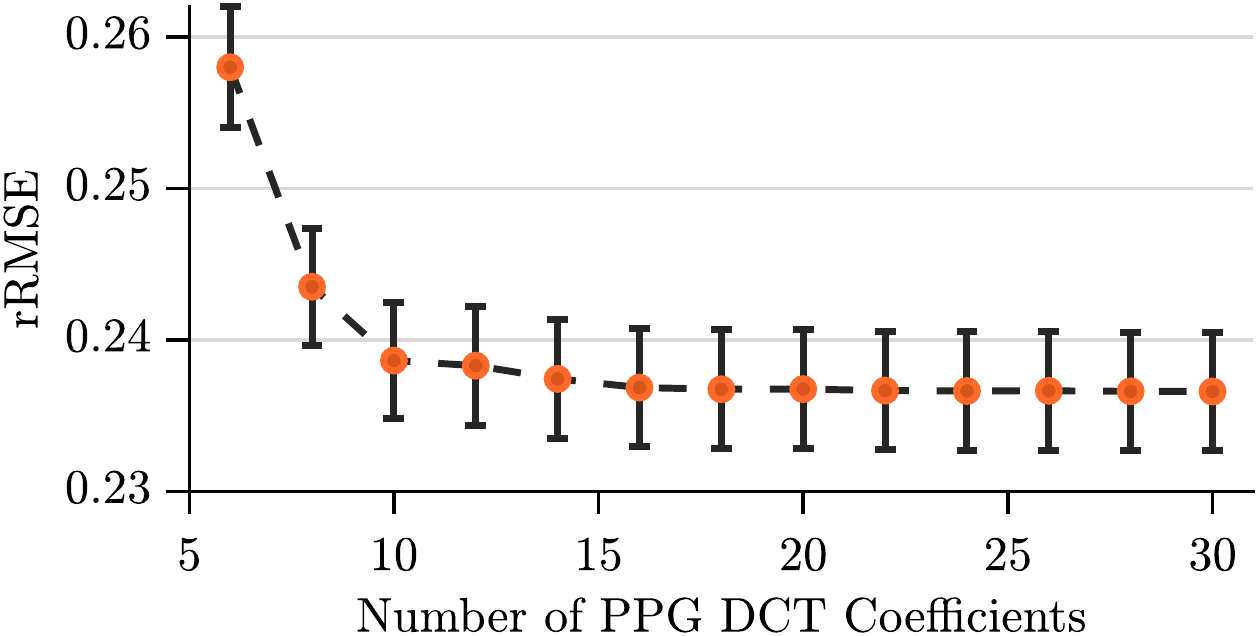}%
\label{subfig::PPGDCTRmse}}
\subfigure[]{\includegraphics[width=0.24\textwidth]{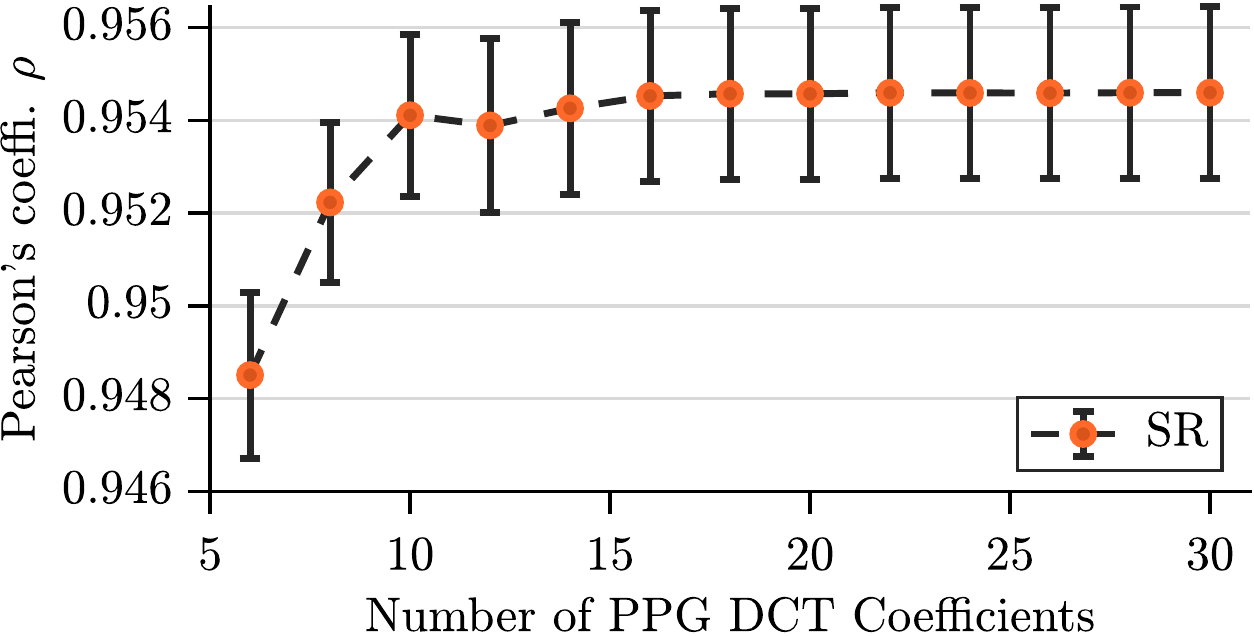}%
\label{subfig::PPGDCTrho}}
\subfigure[]{\includegraphics[width=0.24\textwidth]{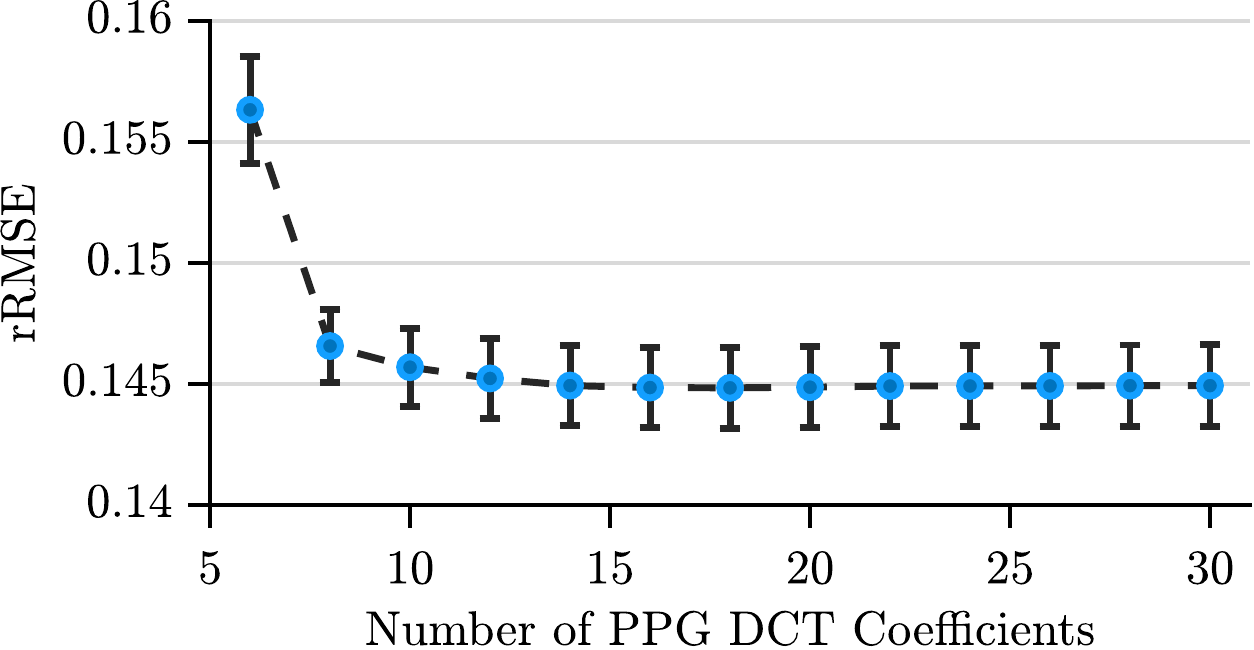}%
\label{subfig::PPGDCTRmse}}
\subfigure[]{\includegraphics[width=0.24\textwidth]{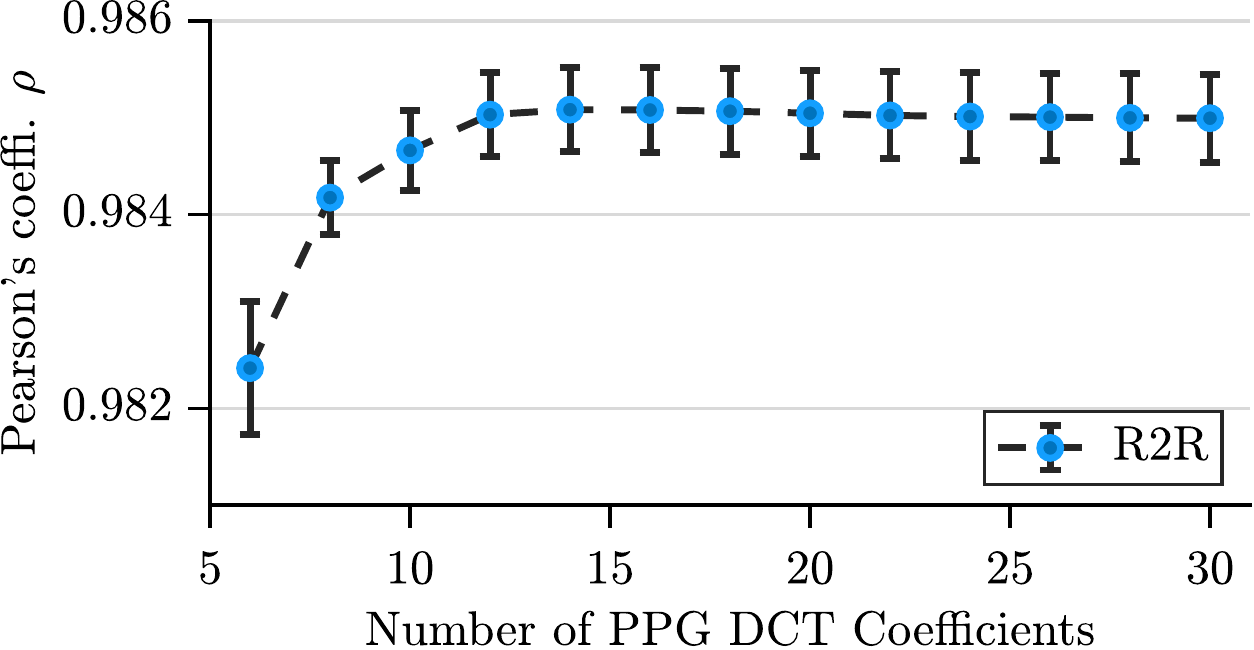}%
\label{subfig::PPGDCTrho}}
\caption{The line plots give the average of r$\textsc{Rmse}$ in (a) and (c) and $\rho$ in (b) and (d) of all sessions in the test set for different numbers of PPG DCT coefficient $m_1$ using SR (a)--(b) and R2R (c)--(d). The vertical bars at each data point shows $3\%$ standard deviation above and below the sample mean.}\label{fig::PPGDCTCoeffi}
\end{figure}

We first cross-validated the number of DCT coefficients of the PPG signal $L_x$ used in the learning system. It is clear that the more variables as predictors, i.e., more PPG DCT coefficients are used in the linear system, the better the performance can be achieved in training. However, we can observe from Fig.~\ref{fig::PPGDCTCoeffi} that the performance of our system in the test set using either SR and R2R becomes saturated as $L_x$ gets larger from 12. This trend of convergence suggests potential model overfitting. $L_x=12$ is thus favorable to us as the system has comparable performance and the model is simpler than those with larger $L_x$.

\begin{figure}[!t]
\centerline{\includegraphics[width=0.48\textwidth]{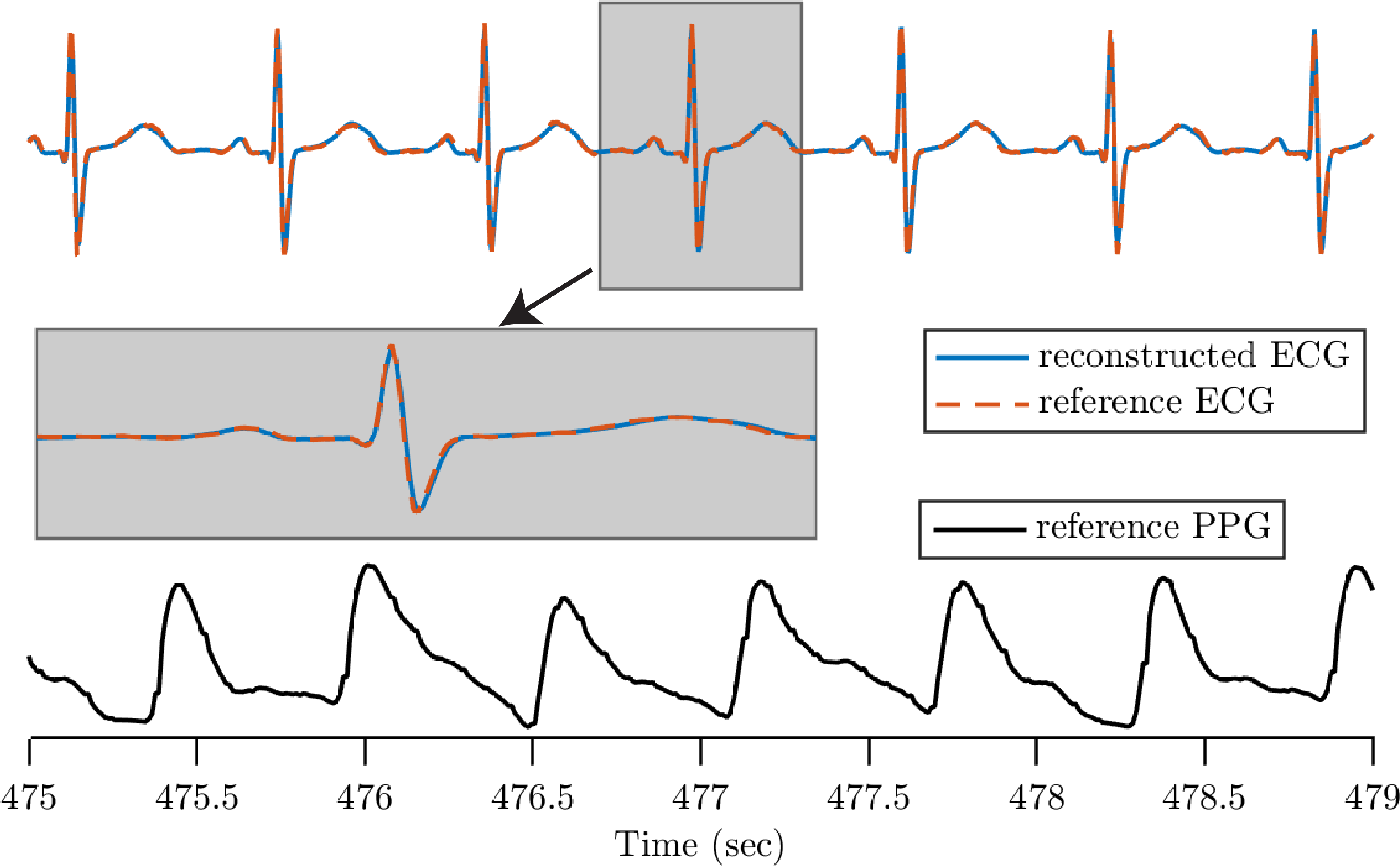}}
\caption{Upper: the reconstructed ECG (blue line) and the reference ECG (red dashed line) waveform of last 4 seconds of the first session (age: 4 years old, weight: 18 kg). Middle: zoomed-in version of the shaded ECG cycle in upper plot. Lower: the corresponding PPG waveform.}
\label{fig::expSysPerformance}
\end{figure}

\begin{table}\centering
\ra{1.3}
\label{tab::cycleseg}
\caption{Sample mean ($\hat{\mu}$) and standard deviation ($\hat{\sigma}$) of $\mathrm{r}\textsc{Rmse}$ and $\rho$ in database using \textsc{R2R} and \textsc{SR} with 12 PPG DCT coefficients.}
\begin{tabular}{@{}ccccc@{}}\toprule
Segmentation Scheme& \multicolumn{2}{c}{r\textsc{Rmse}} & \multicolumn{2}{c}{$\rho$}\\ 
\cmidrule(r){2-3} \cmidrule(l){4-5} 
& $\hat{\mu}$& $\hat{\sigma}$& $\hat{\mu}$& $\hat{\sigma}$\\\midrule
\textsc{SR} & 0.238 &0.118 & 0.954 &0.056\\
\textsc{R2R} & 0.145 & 0.050 & 0.985 &0.013\\\bottomrule
\end{tabular}
\end{table}

We listed the average performance using \textsc{R2R} and \textsc{SR} cycle segmentation schemes in Table~\ref{tab::cycleseg}. The performance is characterized by the sample mean and standard deviation of r\textsc{Rmse} and $\rho$. From the statistics, we learn that overall \textsc{R2R} gives better performance than $\textsc{SR}$ in this dataset.

\par As an example, we show the reconstructed ECG waveform of the last four seconds in the test set of the first subject in Fig.~\ref{fig::expSysPerformance} using the R2R cycle segmentation scheme with $L_x=12$. We can see from the plot that in this case, the system can nearly perfectly reconstruct the ECG and maintain the original shape of the waveform and the location of each PQRST peaks.

\begin{figure}[!t]
\centering
\subfigure[]{\includegraphics[width=0.24\textwidth]{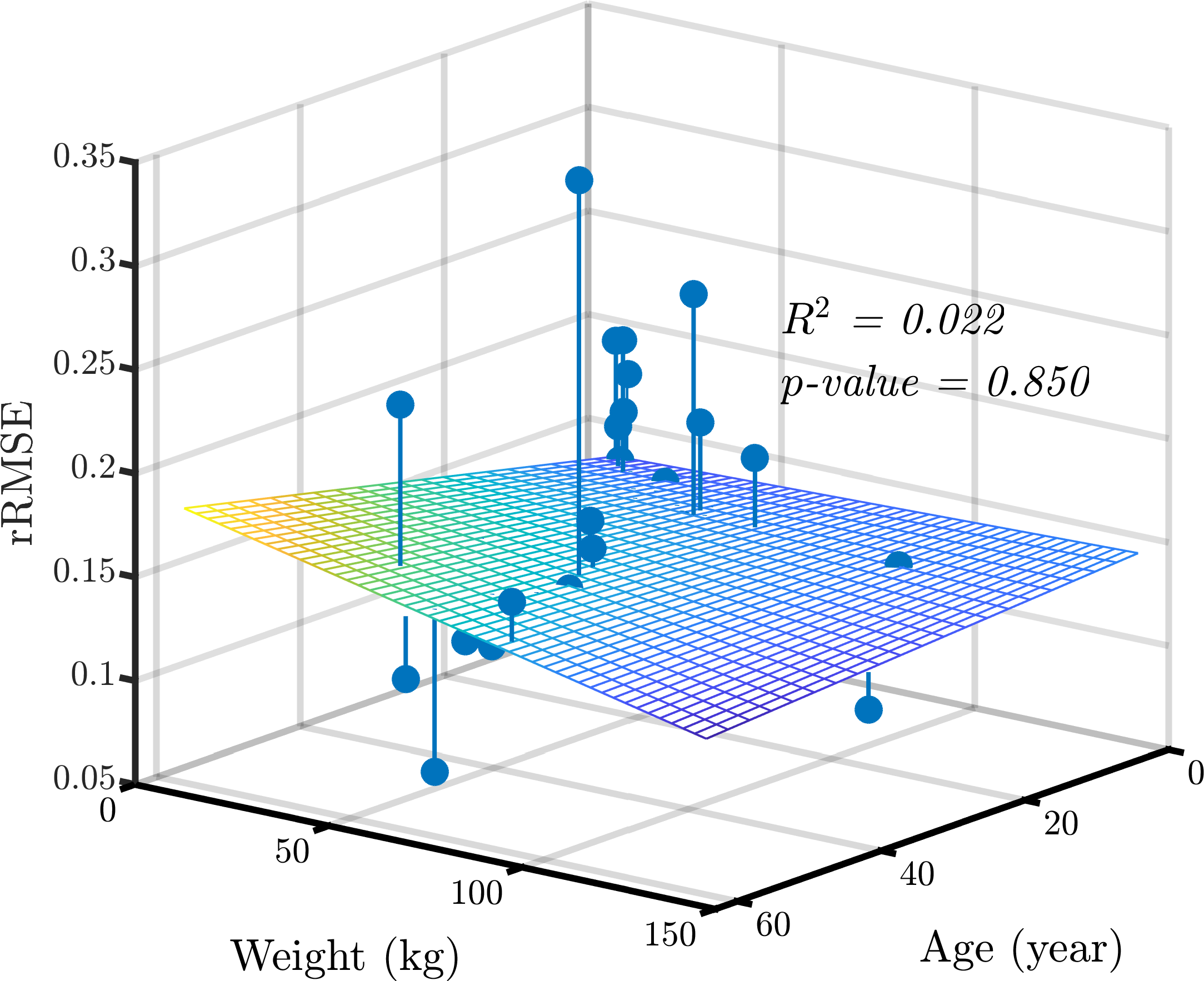}%
\label{subfig::rmsevsage}}
\subfigure[]{\includegraphics[width=0.24\textwidth]{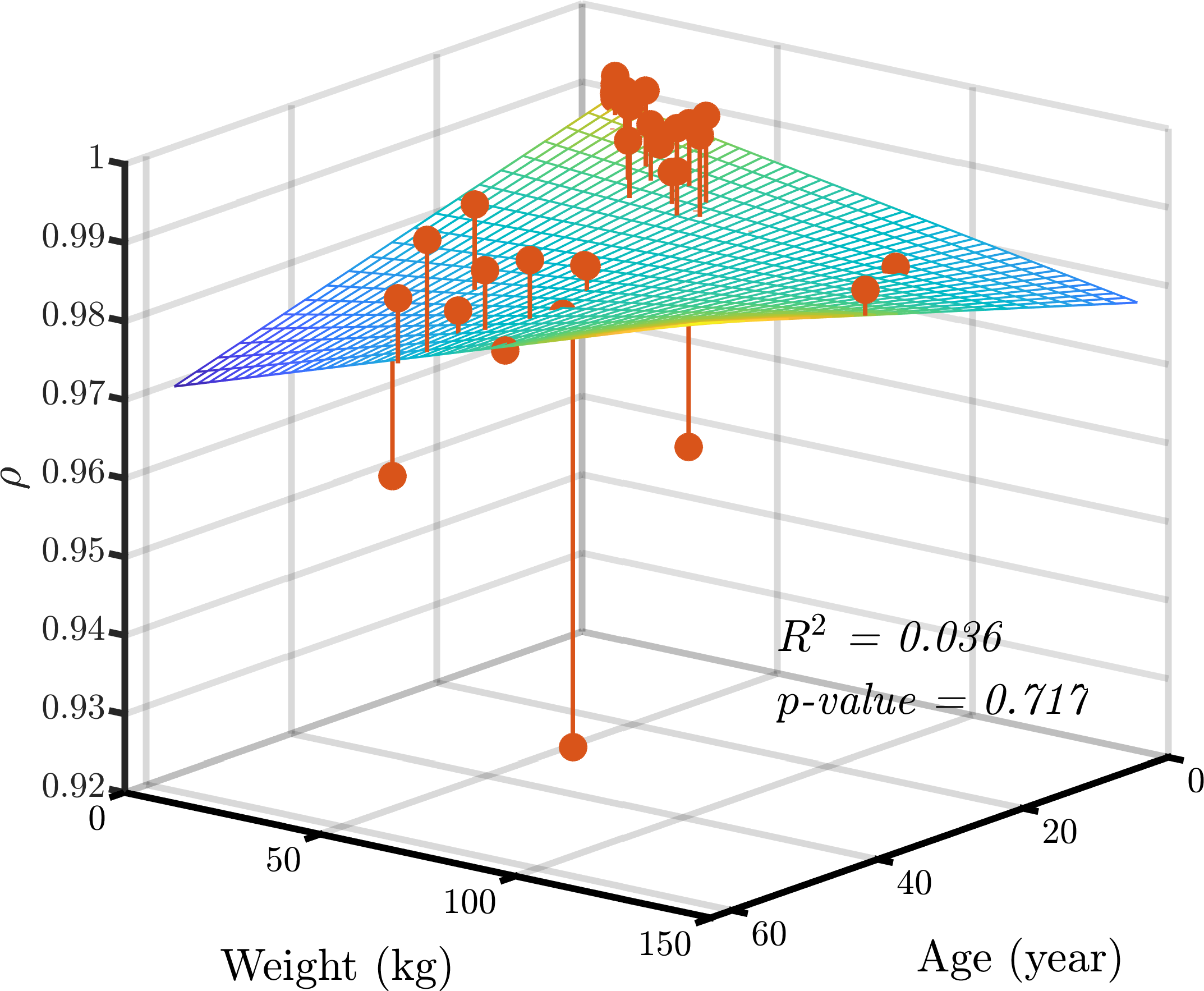}%
\label{subfig::rhovsage}}
\caption{Scatter plots of (a) r\textsc{Rmse} and (b) $\rho$ vs. subjects' weight and age using R2R scheme. Each sample corresponds to one of $42$ sessions. The surface mesh on each plot shows the regressed linear model: $\text{r}\textsc{Rmse}\text{ or }\rho\sim \text{intercept}+\text{age}+\text{weight}+\text{age}\times\text{weight}$. The $R^2$ and the $p$-value of $F$-test is shown on each plot.}\label{fig::errorvsageweight}
\end{figure}

In Fig.~\ref{fig::errorvsageweight}, we plot the r\textsc{Rmse} and $\rho$ of each session with respect to subjects' age and weight respectively in two 3-D plots. We then fitted a linear model with an interaction term for each combination according to the least squares criterion. An $F$-test is performed to test whether subjects' profile, i.e., age and weight, can significantly affect the performance of the algorithm in each metric. $F$-tests results of high $p$-values shown in Fig.~\ref{fig::errorvsageweight} reveal that the performance of the algorithm is not dependent on age and weight.

\section{Conclusion}\label{sec:conclusion}
This paper presents a learning-based approach to reconstruct ECG signal from PPG. The algorithm is successfully evaluated in a subject-dependent fashion on a widely-adopted database. We cross-validate the system hyper-parameters and justify the algorithm's accuracy and consistency. As a pilot study, this work demonstrates that with a signal processing and learning system that is justified in each design step, we are able to precisely reconstruct ECG signal by exploiting the relation of the two measurements.

\bibliographystyle{IEEEtran}

\bibliography{PPG2ECG.bib}

\end{document}